\documentclass{article}

    \PassOptionsToPackage{numbers, compress}{natbib}

    \usepackage[preprint]{neurips_2024}

\usepackage[utf8]{inputenc} 
\usepackage[T1]{fontenc}    
\usepackage{hyperref}       
\usepackage{url}            
\usepackage{booktabs}       
\usepackage{amsfonts}       
\usepackage{nicefrac}       
\usepackage{microtype}      
\usepackage{xcolor}         

\usepackage[numbers]{natbib}
\usepackage[hypcap=false]{caption} 

\usepackage{hyperref} 
\usepackage{listings} 
\usepackage{graphicx}
\usepackage{subfig}
\usepackage{wrapfig}
\usepackage{multirow}
\usepackage{caption}

\usepackage{tabularx} 
\usepackage{geometry} 
\usepackage{verbatimbox} 
\usepackage{cprotect}

\definecolor{codegreen}{rgb}{0,0.6,0}
\definecolor{codegray}{rgb}{0.5,0.5,0.5}
\definecolor{codepurple}{rgb}{0.58,0,0.82}
\definecolor{backcolour}{rgb}{0.95,0.95,0.92}
\definecolor{amber}{rgb}{1.0, 0.75, 0.0}

\title{DafnyBench: A Benchmark for Formal Software Verification}

\author{
  Chloe Loughridge$^*$ \\
  Harvard College \\
  \texttt{cloughridge@college.harvard.edu} \\
  \And
  Qinyi Sun$^*$$^{\dagger}$ \\
  Massachusetts Institute of Technology \\
  \texttt{wendysun@mit.edu} \\
  \And
  Seth Ahrenbach \\
  \texttt{seth.ahrenbach@omnifederal.com} \\
  \And
  Federico Cassano \\
  Northeastern University \\
  \texttt{cassano.f@northeastern.edu} \\
  \And
  Chuyue Sun \\
  Stanford University \\
  \texttt{chuyues@stanford.edu} \\
  \And
  Ying Sheng \\
  Stanford University \\
  \texttt{ying1123@stanford.edu} \\
  \And
  Anish Mudide \\
  Massachusetts Institute of Technology \\
  \texttt{amudide@mit.edu} \\
  \And
  Md Rakib Hossain Misu \\
  University of California Irvine \\
  \texttt{mdrh@uci.edu} \\
  \And
  Nada Amin \\
  Harvard University \\
  \texttt{namin@seas.harvard.edu} \\
  \And
  Max Tegmark \\
  Massachusetts Institute of Technology \\
  \texttt{tegmark@mit.edu}
}

\begin{document}

\maketitle

\def\thefootnote{*}\footnotetext{Equal contribution. Order determined alphabetically.}
\def\thefootnote{\textdagger}\footnotetext{Corresponding author.}
\def\thefootnote{\arabic{footnote}}

\begin{abstract}
We introduce DafnyBench, the largest benchmark of its kind for training and evaluating machine learning systems for formal software verification.
We test the ability of LLMs such as GPT-4 and Claude 3 to auto-generate enough hints for the Dafny formal verification engine to successfully verify over 750 programs with about 53,000 lines of code. The best model and prompting scheme achieved 68\% success rate, and we quantify how this rate improves when retrying with error message feedback and how it deteriorates with the amount of required code and hints. We hope that DafnyBench will enable rapid improvements from this baseline as LLMs and verification techniques grow in quality.
\end{abstract}

\section{Introduction}
\label{intro}

Rapidly improving Large Language Models (LLMs) ~\cite{bubeck2023sparks,claude_3_tech_report,gemini_tech_report} 
are helping accelerate software development through co-pilots and other program synthesis tools. 
But how can we ensure that LLM-generated code meets our specifications and reliably does precisely what it is supposed to do?
Indeed, this remains a persistent problem even with human-written code: 
major code-testing efforts failed to prevent e.g. bugs causing an Ariane-V rocket explosion \cite{esa2019}
and embarrassing security vulnerabilities 
in ssh \cite{heartbleed2014} 
and the  Bash shell  \cite{shellshock2014}. The latter was built into the Unix operating system for 25 years before being discovered.

Although {\it formal verification} can guarantee perfect reliability, providing rigorous mathematical proof that software meets specification, it has yet to gain widespread adoption because it is costly. 
Formally verifying code can easily take more than ten times as much human work as writing it in the first place. 
Moreover, existing formal-verification tools tend to involve a major learning curve above and beyond just learning to code, 
greatly reducing the pool of people able to do this work.

The premise of this paper is that AI will soon be able to greatly facilitate formal verification, and hopefully even fully automate it one day.
This would drive its cost to near-zero, dramatically increase its adoption and dramatically reduce the prevalence of buggy software.
It is easy to imagine formal verification becoming simply a built-in final step of future compilers, which discover code problems and
perhaps even fix them automatically.
This optimistic premise is based on the close analogy with automated theorem proving, where AI produces formal proofs not about code but about mathematical theorems. Fueled by the advent of  benchmarks totaling over 100,000 theorems,
AI tools have during the past few years improved their proof success fraction to over 82\% \citep{Polu2020Sep,Lample2022May}.

Unfortunately, formal verification sorely lacks correspondingly large benchmarks: the largest of their kind are
{\it  Clover} \citep{sun2024clover} and {\it dafny-synthesis} \citep{MRHMisuDafnyFSE24},  containing 66 and 153 programs, respectively. 
There is room for expanding not only their size, but also their level of difficulty:  
For example, {\it  Clover} is limited to single-function programs, and sometimes the formal specification for the program directly 
repeats the implementation of the algorithm (see Appendix \ref{over_detailed_spec}).
To support automation of formal verification, the goal of the present paper is to provide such a benchmark expansion.
We do so by assembling a suite of formally verified programs written in {\it Dafny}, a formal verification language that was developed for easy adoption by programmers due to its similarity with popular imperative programming languages such as Python and C++
\cite{leino2023program}.
In order for formal verification to succeed, most of these programs require supplementary text constituting ``hints'' to the automated theorem prover.

The rest of this paper is organized as follows. We summarize related work in Section \ref{related_work}, describe our benchmark construction in Section \ref{dataset_construction}, and quantify the ability of current LLMs to solve benchmark verification tasks in Section \ref{experiments}. We summarize our results and discuss promising opportunities for further work in Section~\ref{discussion} . We provide further details on the benchmark construction and evaluation in appendices.

\section{Related Work}
\label{related_work}

As summarized in Table \ref{related_work_table} below, there is a striking lack of training data for formal verification: while there are hundreds of thousands of training examples for proving mathematical theorems and 
over ten thousand training examples for 
synthesizing programs, there are  
only $66+153=219$ for proving program correctness.
This motivates our work in the current paper to expand the benchmarks from \textit{Clover} and \textit{dafny-synthesis}.

\begin{table}[ht]
    \caption{Summary of popular machine-learning benchmark datasets for proving mathematical theorems, synthesizing programs, and formally verifying programs.
    Size is measured by the number of samples in each dataset. In the formal reasoning datasets, each sample is usually a math problem or a theorem. In the program synthesis and verified software programming benchmarks, each sample corresponds to a program. \\}
    \centering
    \begin{tabular}{lll}
        \toprule
        \textbf{Category} & \textbf{Dataset} & \textbf{Size} \\
        \midrule
        \multirow{5}{*}{\textbf{Mathematical theorem proving }} 
        & CoqGym \citep{yang2019} & 71,000 proofs \\
        & LeanDojo \citep{leandojo2023} & 98,734 proofs \\
        & PISA \citep{pisa2021} & 138,000 proofs \\
        & Natural Proofs \citep{naturalproofs2022} & 15,000 proofs \\
        & Archive of Formal Proofs \citep{klein2014} & 1 million lines of code \\        
        \midrule
        \multirow{5}{*}{\textbf{Unverified program synthesis}}
        & APPS \citep{hendrycks2021measuring} & 10,000 programs \\
        & HumanEvalX \citep{zheng2023codegeex, chen2021evaluating} & 165 programs \\
        & MBPP \citep{austin2021program} & 974 programs \\
        & SWEBench \citep{jimenez2023swebench} & 2,294 programs \\
        & LiveCodeBench \citep{jain2024livecodebench} & grows weekly \\
        \midrule
        \multirow{2}{*}{\textbf{Formal software verification}} & Clover \citep{sun2024clover} & 66 programs \\
        & Dafny-synthesis \citep{MRHMisuDafnyFSE24} & 153 programs \\
        \bottomrule \\
    \end{tabular}
    \label{related_work_table}
\end{table}

The 66 programs in the \textit{Clover} benchmark are human-written. In contrast, \textit{dafny-synthesis} translates 153 MBPP problems from Python to Dafny using GPT-4. While this method is more efficient than manual translation, it could potentially skew the distribution of represented problems away from real-world Dafny problems that may be too hard for GPT-4 to verify on its own \citep{MRHMisuDafnyFSE24}. Our dataset counterbalances this potentially skewed distribution by introducing problems verified by human programmers on GitHub.

\textit{Clover} proposes the most sophisticated benchmark evaluation strategy to date for formally verifiable software: the authors suggest a six-way consistency check between code, docstrings, and hints. Their checker achieves an 87\% acceptance rate of correct implementations on the \textit{Clover} benchmark while rejecting all incorrect implementations \citep{sun2024clover}. The authors note that equivalence checking with natural language is currently weak, but can hopefully be improved upon \citep{sun2024clover}. We do not yet implement the full \textit{Clover} evaluation scheme in DafnyBench, and instead deem a benchmark program "solved" if a model can make it pass the Dafny verifier without modifying the \texttt{requires} and \texttt{ensures} statements in the program and without using \texttt{\{:verify false\}} or \texttt{assume false} (see Appendix \ref{verification-examples} for further details).

\section{DafnyBench Dataset Construction}
\label{dataset_construction}

\subsection{Sourcing Ground Truth Programs}

In total, our DafnyBench benchmark contains 782 \verb+ground_truth+ stand-alone Dafny programs that compile. These problems come from the following sources:

\begin{itemize}
    \item \textbf{GitHub Scrape}: We scraped all publicly available Dafny files on GitHub published on the before the end of 2023. The relevant files were returned from the GitHub API using the \texttt{language:Dafny} search command. We then de-duplicated these files using a minhash de-duplication script written by Chenghao Mou (described in Appendix \ref{appendixA}). The de-duplication process reduced the number of \texttt{.dfy} files from $\sim$15,000 to $\sim$5,000. We then attempted to verify each of these remaining files using the \texttt{dafny verify} command with a local installation of Dafny 4.3.0, and removed any files that did not verify. At this stage, we removed all of the files from the \textit{Clover} repository \cite{sun2024clover}, which had already been formatted as benchmark files. This left 1,112 files. We found that 374 of these files lacked \textit{ensures} statements, and 459 of lacked \verb+assert+ and \verb+invariant+ clauses. We removed the union of these sets, which left us with 556 \verb+ground_truth+ files. Out of these files, 113 verify without any compiler hints. To mitigate data contamination, models run on our benchmark should ideally not be trained on data from the repositories listed in Appendix \ref{appendixD}.
    \item \textbf{Clover}: We added 62 ground truth textbook Dafny programs provided by the \textit{Clover} dataset \citep{sun2024clover}. We formatted these to fit our benchmark style and removed their compiler hints. Out of these files, 23 verify without any compiler hints.
    \item \textbf{Dafny-synthesis}: Finally, we included 164 Dafny programs provided by the \textit{dafny-synthesis} benchmark. These problems have been translated from the MBPP benchmark \citep{MRHMisuDafnyFSE24}. Out of these files, 72 verify without any compiler hints.

\end{itemize}

The \verb+ground_truth+ programs in our dataset have on average 2.12 methods, 1.03 functions, and 1.40 lemmas. This places the mean complexity of our examples at a level higher than \textit{Clover} alone, which has only one stand-alone method per example.

\begin{table}[ht]
    \caption{Mean and maximum values that describe attributes of a DafnyBench test program.}
    \label{metadata_summary}
    \centering
    \begin{tabular}{lll}
        \\
        \toprule
        & Mean & Max \\
        \midrule
        \# Methods & $2.12$ & $42$\\
        \# Functions & $1.03$ & $42$\\
        \# Lemmas & $1.40$ & $35$\\
        \# Characters & $1916.47$ & $28736$\\
        \# Hint characters & $261.23$ & $6019$\\
        \bottomrule \\
    \end{tabular}
\end{table}

\subsection{Task Design: Fill Hints}

We have fully implemented the \verb+fill_hints+ task. For this task, we took a \verb+ground_truth+ program, removed all of its hints (i.e., all of the \verb+assert+ and \verb+invariant+ statements in the body of the code), and asked LLM to fill hints back in so that the resulting program could be verified with Dafny.

\begin{figure}[ht!]
    \centering
    \includegraphics[width=0.95\linewidth]{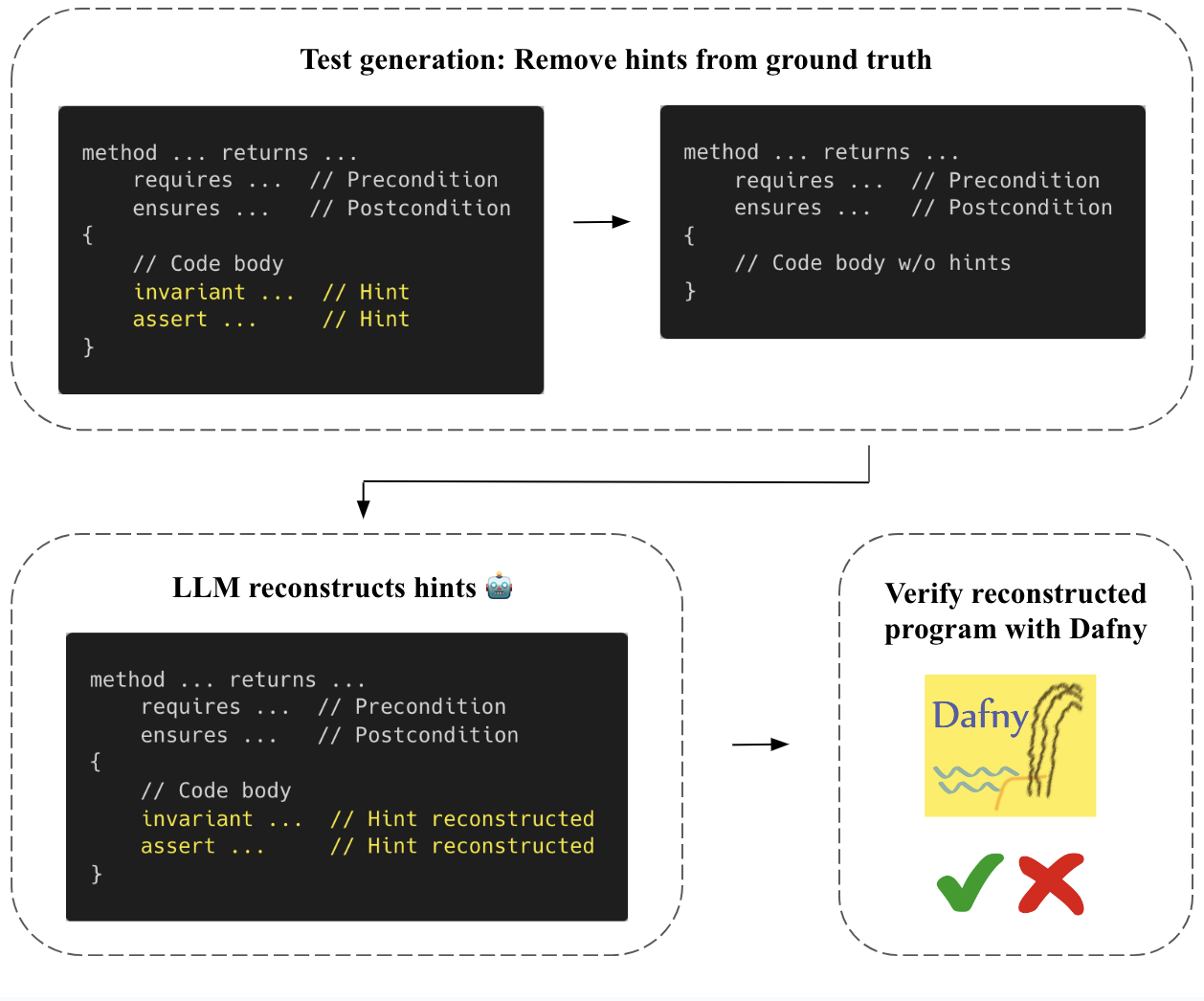}
    \caption{Overview of evaluating LLM on a DafnyBench test program.\\ \hspace{1cm}}
    \label{fig1}
\end{figure}

We do not demarcate from where these hints have been removed, i.e., we do not insert \texttt{/* TODO */} after we remove each annotation, which would make the task easier and not reflective of models utility in real-world use cases.

\paragraph{Evaluation Metric} An LLM's attempt to fill hints back in for a test program is counted as a success if all following conditions are satisfied: 1) The reconstructed program is verified with Dafny; 2) LLM preserves all preconditions (\verb+requires+ statements) and postconditions (\verb+ensures+ statements); and 3) LLM does not use \verb+{:verify false}+ or \verb+{assume false}+ to "cheat."

\begin{figure}[ht!]
\begin{lstlisting}
method LinearSearch<T>(a: array<T>, P: T -> bool) returns (n: int)
    ensures 0 <= n <= a.Length
    ensures n == a.Length || P(a[n])
    ensures forall i :: 0 <= i < n ==> !P(a[i])
{
    n := 0;
    while n != a.Length
        invariant 0 <= n <= a.Length
        invariant forall i :: 0 <= i < n ==> !P(a[i])
    {
        if P(a[n]) {
            return;
        }
        n := n + 1;
    }
}
\end{lstlisting}
\caption{An example \texttt{ground\_truth} program that is fully verified with Dafny. To create the \texttt{fill\_hints} task, we would remove the \texttt{invariant} lines from the program above.}
\label{fig2}
\end{figure}

\section{Experiments}
\label{experiments}

In this section, we report success rates
for different models on the \verb+fill_hints+ task, as well as provide some insight into current LLMs' capabilities at writing hints for formal verification.

\subsection{Prompts \& Hyperparameters}

We tried to keep prompts and hyperparameters mostly the same across models in order to reduce the difference between model performances that is caused by hyperparameters. However, the prompts are not fully identical. For example, when we ask LLM to simply return the hints-filled program without any explanation, Claude 3 tends to add explanations that interfere with Dafny compilation. Thus, we had to adjust some prompts slightly to fit each model's peculiarities.

For hyperparameters, we set \verb+max_tokens+ $= 4096$, which corresponds to the lowest max output token limit among all the evaluated models, and we set \verb+temperature+ $= 0.3$. We gave each model up to $n=10$ attempts at a given file. If it succeeded on an attempt before the $n^{\rm th}$, it would be early stopped. If the model failed on any of the intermediate attempts, it received the Dafny error message and was asked to filled in the hints again with the error message taken into consideration. If it failed on all $n$ attempts, it was considered to fail on that specific test program.

\subsection{Basic Results}

We tested GPT-4o, GPT-4 Turbo \cite{openai_2023_gpt4}, GPT-3.5 Turbo \cite{brown2020gpt3}, Claude 3 Opus \cite{claude_3_tech_report}, and CodeLlama-7b-Instruct-hf \cite{huggingface2022llama} on the $782$-program benchmark. 
Table \ref{model_stats} shows that Claude 3 Opus performed best, achieving a success rate $\sim 68\%$.

\subsection{Difficulty Utilizing Dafny Error Messages} 
Figure \ref{fig3} shows how the cumulative success rate improved with more attempts $n$. We see that the best models succeeded on the first try about 54\%, with rapidly diminishing returns after that, approaching a plateau about 65\% for $n \sim 5$. This suggests that the LLMs are not great at taking Dafny error messages into consideration, or struggle to cope with the underlying task.

\begin{minipage}{\textwidth}
    \begin{minipage}[b]{0.45\textwidth}
        \centering
        \begin{tabular}{lll}\hline
        \toprule
        Model & \% Success \\
        \midrule
        No LLM & $26.9$ \\
        GPT-3.5 Turbo & $44.0 \pm 1.8$ \\
        GPT-4 Turbo & $59.8 \pm 1.8$ \\
        GPT-4o & $59.3 \pm 1.8$ \\
        Claude 3 Opus & \textbf{67.8} $\pm 1.7$ \\
        CodeLlama-7b-Instruct-hf & $28.0 \pm 1.6$ \\
        \bottomrule
        \end{tabular}
        \captionof{table}{
        Models' success rates at writing formally verifiable hints for DafnyBench, with $n = 10$ attempts given. Dafny succeeds in auto-verifying some programs even without hints, corresponding to the ``No LLM"  $26.9\%$ success rate baseline.\\}
        \label{model_stats}
    \end{minipage}
    \hfill
    \begin{minipage}[b]{0.46\textwidth}
        \centering
        \includegraphics[width=\linewidth]{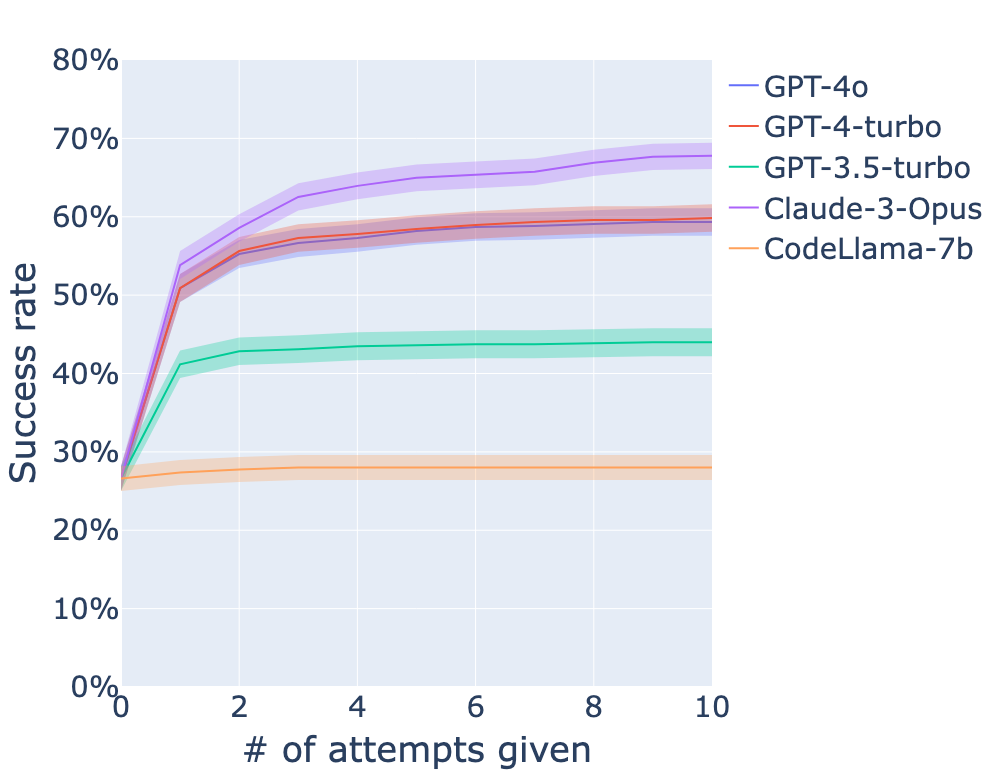}
        \captionof{figure}{Success rate vs. number of\\attempts given.\\}
        \label{fig3}
    \end{minipage}
\end{minipage}

\subsection{Difficulty Grows with Program Size}

Figure \ref{fig:program_length} show that the success rate drops with program size. 
An obvious explanation could be that there is more to verify and more hints needed. Also, 
as a program gets longer, there may be more dependencies among variables, functions, methods, and classes, increasing the overall verification difficulty level.

\subsection{Difficulty Grows with Hint Quantity}

\begin{figure}[ht]
  \subfloat[]{
	\begin{minipage}[c][0.6\width]{
	   0.49\textwidth}
	   \centering
	   \includegraphics[width=1\textwidth]{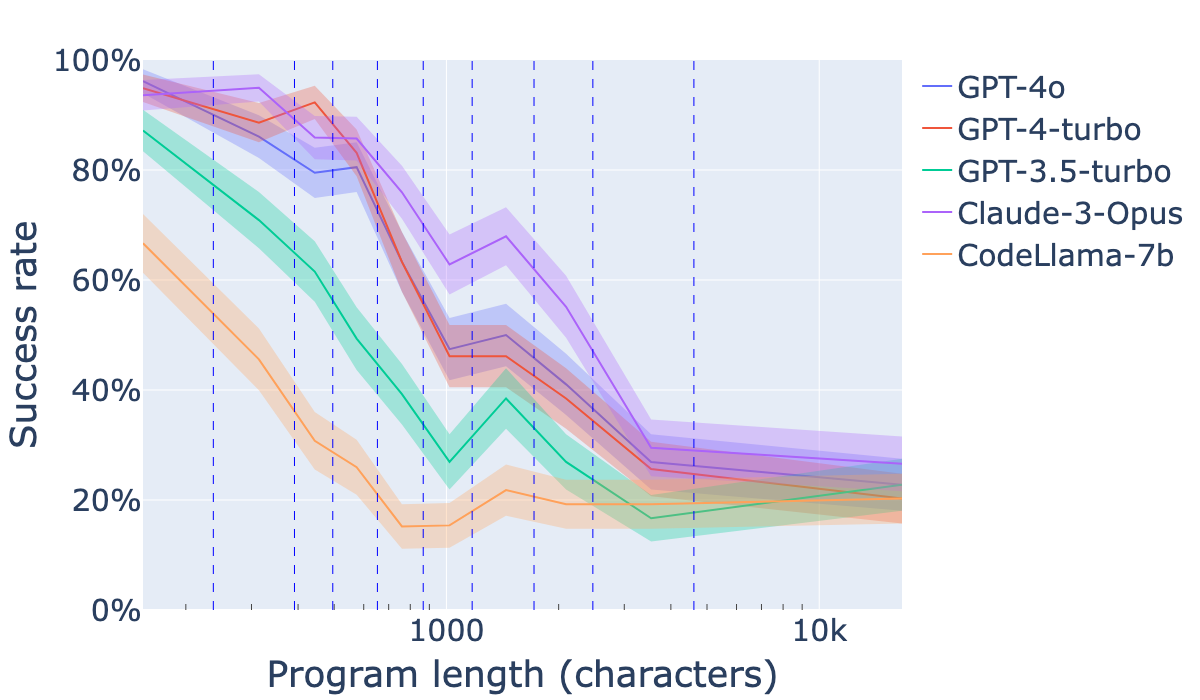}
        \label{fig:program_length}
	\end{minipage}}
 \hfill 	
  \subfloat[]{
	\begin{minipage}[c][0.6\width]{
	   0.49\textwidth}
	   \centering
	   \includegraphics[width=1\textwidth]{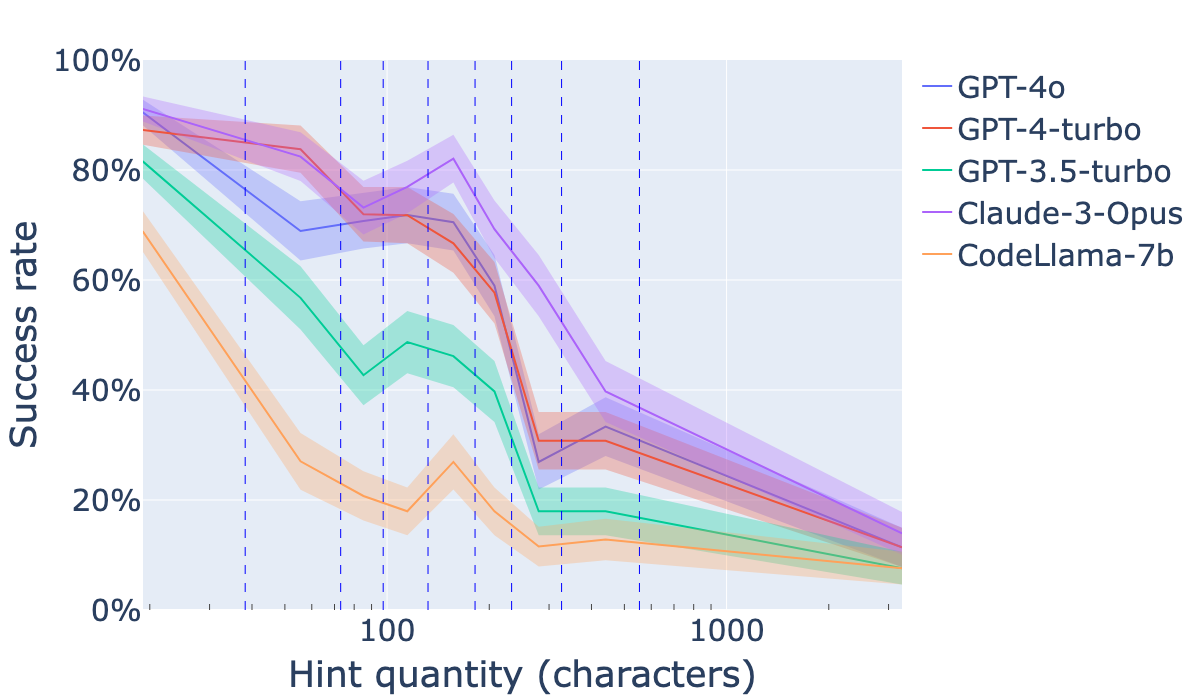}
        \label{fig:hint_quantity}
	\end{minipage}}
\caption{\textbf{Mean success rate of each bin vs. program length (a)}, and \textbf{mean success rate of each bin vs. hint quantity (b)}. The vertical lines indicate the bin boundaries used, where the bins have an almost uniform distribution of the programs. Note that the bins are different for the two metrics. For better visual clarity, the scales are adjusted for both plots and their $x$-axes do not start at $0$ character.}
\end{figure}

Figure \ref{fig:hint_quantity} shows that the success rate drops with the hint quantity, defined as the number of characters in the lines of compiler hints. In other words, the success rate drops with the amount of work that the LLM needs to do (the amount of text that it needs to insert in the right places).

\section{Discussion and Conclusions}
\label{discussion}

We have assembled the largest machine-learning benchmark to date 
for formal software verification and made it publicly 
available on GitHub at \url{https://github.com/sun-wendy/DafnyBench}.
We also tested five large language models on this benchmark, including one open source model.

We found that Claude 3 Opus achieved $\sim 68\%$ accuracy on our benchmark, with even better success on programs that were shorter or involved less hint text than the benchmark average. GPT-4 Turbo came second with $\sim 60\%$ accuracy. Meanwhile, CodeLlama-7b-Instruct-hf only achieved a marginal improvement in accuracy compared to our "No LLM" baseline. While in certain cases it succeeds in copying and lightly modifying programs that already verify without compiler hints, it fails to add compiler hints to programs that don't verify without them.

\subsection{Opportunities for Larger Benchmarks}
It will be valuable to further expand formal verification benchmarks, 
which still remain more than two orders of magnitude smaller than corresponding 
benchmarks for mathematical theorem proving. 
One convenient way to expand the number of available problems may involve incorporating Dafny programs from GitHub that have dependencies spread across multiple files (while \textit{DafnyBench} encompasses increasingly complex multi-step programs, its programs each fit in a single file, avoiding the intricacies associated with distributed files or the integration of external libraries).

Perhaps models that perform especially well on this initial benchmark can later be used to expand it by translating existing Python benchmark problems into Dafny, Rust \cite{RustBook} or other popular formal verification languages.

A subset of the programs we scraped from GitHub do not have appropriate docstrings. By building a benchmark with better code documentation, models may be able to leverage helpful contextual information to better constructing verification hints. 

\subsection{Benchmark Evaluation Limitations} 
Data contamination emerges as a potentially significant limitation for evaluating LLMs on this benchmark. Scraping data from platforms such as GitHub introduces risks of leveraging previous models' training data into the benchmark evaluation, potentially artificially inflating the abilities of certain models.

Another limitation emerges in that this benchmark does not assess a model's competence in translating natural language into concise formal specifications. Arguably, this conversion is a  demanding and crucial skill we seek from language models: the capacity to validate, beyond merely verifying code. The pivotal question is whether a model can assist in identifying the essential properties an algorithm must fulfill. Currently, evaluating this ability presents significant challenges. The \textit{Clover} paper stands as a prominent example in this area, highlighting the complexity of translating natural language descriptions into formal specifications that can be effectively used for validation. This provides an exciting frontier for future work, which we begin to brainstorm in Appendix \ref{appendixC}.

\subsection{Opportunities for Improved LLM Results}

It will be interesting to test this benchmark on additional LLMs, both existing ones such as Gemini \cite{gemini_tech_report} and Grok \cite{grok1_huggingface} and upcoming ones.
Furthermore, we evaluated the models with a fixed temperature setting and a max output token limit of 4096, and we used prompts that were manually but not very systematically tuned for effectiveness (see Appendix \ref{appendixB}) --- all of these choices probably leave room for improvement.

We do not yet provide an official training dataset or models custom-trained to do well on the DafnyBench evaluation set. However, we do provide the full json file produced by the GitHub scrape, and we separately provide the names of the files we use for the evaluation benchmark. Hence it is possible for researchers to use files from the Github scrape that are not used in the benchmark as training data, though we cannot at this time provide strong guarantees on similarity between such training problems and the benchmark problems. Pre-training on this type of data may boost large language model performance on DafnyBench.

We also see great opportunity for LLM-related innovation on the algorithmic side: out-of-the-box LLMs provide a floor but not a ceiling for possible performance on this benchmark. For example, fine-tuning or search-based inference-time algorithms might boost models' performances on this benchmark \citep{brandfonbrener2023verified}.

\subsection{The Promise of Better LLM-Powered Verifiers}

LLMs also have potential to improve formal verification in more profound ways than mentioned above, when used in combination with other AI tools.
For example, they can help automate the identification of sub-goals and hints,  exponentially reducing the search space for automated theorem provers and SAT solvers. A good software developer is likely able to specify the high level assurance properties of a piece of code. However, in trying to prove that the given code satisfies these high level properties, numerous, sub-goals must be identified, proven, and leveraged correctly in the broader context. Software developers often lack familiarity with the complexities of proof sub-goals and hints. LLMs offer a way to bridge this gap between software developers and formal verification. 

Achieving this requires benchmarks suitable for improving the performance and generality of LLMs with respect to software verification. Bigger, more general benchmarks can be used to train LLMs to specify sub-goals and hints in formats most useful to the presently available provers and solvers. Benchmarks covering broad ground, from cryptography, lambda calculus, embedded systems, and avionics, in a variety of widely used programming languages suitable for verification, will help create LLMs that can take real-world software, automatically process and serve it to verification tools, and inform the developer in near real time about the correctness of the code. The problem is analogous to that solved by existing automated theorem provers and model checkers in the domain of mathematics. They address the problem, when given a set of constraint formulas or background theorems, whether a candidate formula is satisfiable or derivable. Many clever algorithms have increased the degree of automation available to mathematical theorem proving over time. LLMs should be able to help similarly improve  automation for software verification. For a survey on the application of deep learning to automated theorem proving, see \cite{li2024dl4tp}. 

In order to formally specify a correctness property for a programming language, some formalization of the lower level language's semantics must be represented in a higher level specification language. A lower level language with well-defined semantics to begin with makes this easier. For languages lacking well-defined semantics, such as C, JavaScript, and Python, a well-defined subset may suffice \cite{Blazy-Leroy-Clight-09}. Programming languages fall on a spectrum of well-defined semantics, with higher level languages like Haskell on one end, and C on the other. Rust falls in a particularly nice intermediate place, with a strongly typed, functional semantics and macros for achieving side effects. An ecosystem of formal verification tools has begun to emerge for Rust, due to its nice semantics and popularity as a practical programming language 
\cite{matsakis2014rust}. A benchmark leveraging this ecosystem for LLMs would likely compound on this progress dramatically. Multiple formal verification tools compile to Rust or extract correct Rust code. For example, Dafny can compile to Rust, and other tools for extracting Rust from Coq exist \cite{Coq-refman, annenkov_milo_nielsen_spitters_2022}. In this case, Rust would be considered the low level language, and Dafny and Coq would serve as candidate specification languages. A workflow might be possible such that a developer working in Rust could have a LLM assistant that identifies correctness properties for the code, either automatically or provided at a high level by the developer, and produces appropriate artifacts for verifying correctness via multiple tools for improved assurance.


\subsection{The Promise of Auto-Verifying Program Synthesis}


Above we discussed the challenge of verifying existing pre-programs. 
Anther promising approach is use program-synthesis techniques that produce not only programs but also proofs of their correctness, all at the same time. This makes intuitive sense, since when a human programmers writes code, they typically have an informal proof in their head for why this code is correct.

In other words, in addition to bridging the gap from low level implementation to high level specification in the upward direction, LLMs can offer assistance in generating provably correct low level code from high level specifications via program synthesis. Current approaches to program synthesis enable engineers to encode a desired specification in a high level language, and then through a (hopefully) verified correct compiler generate correct low level code in a language like VHDL \cite{VHDL} or Verilog \cite{Verilog} for hardware synthesis. Indeed, the compilation of Dafny code to Rust or Python is an example of program synthesis. Program synthesis is limited by the need for a special purpose language or compiler to be constructed and verified correct in its own right. For example, ReWire is a domain specific language defined as a subset of Haskell \cite{Procter2015}. Using ReWire, engineers can specify hardware properties and then through the Haskell compiler, synthesize VHDL that is guaranteed to satisfy the specifications. ReWire itself was manually verified correct using the Coq Interactive Theorem Prover. In order to add a new high to low path, a new language or compiler must be defined and verified. If an engineer needs to synthesize correct Verilog rather than VHDL, they must first learn Caisson \cite{Li2011}.

LLMs offer a way to generalize this approach. Starting with a high level language, an engineer might be able to specify a system and then leverage a LLM to generate low level code with the corresponding loop invariants, weakest pre-conditions, strongest post-conditions, etc, included. In the limit, an engineer might be using a natural language to describe the system and its desired assurance properties, with the LLM performing translation, annotation, and even suggesting additional correctness properties. Early results indicate that an LLM that is able to converse with a human when producing a program can reduce the error rate against a simple programming benchmark by half \cite{austin2021program}. If instead of receiving feedback from a human, the LLM were to interact with a suite of formal verification tools, we expect further improvements. We could avoid hallucination problems by relying on the LLM to generate the code and formal specification, but relying on an established verification tool to perform the model checking or proof verification itself. The LLM's translation process need not itself be verified, because it can try multiple times to produce a verifiable output. The LLM must be capable of generating code that is appropriately annotated for theorem proving, which is exactly the skill assessed by test benches like that described here. The more theorem proving tools and programming languages that LLMs are trained and assessed on, the more auto-verifying program synthesis options become available. To return to the previous example, a LLM proficient at ReWire, Caisson, and myriad other software verification techniques, might be given a ReWire specification as input and told to produce correct Verilog as output. The ReWire specification contains the high level correctness properties that must be satisfied. The task is to synthesize Verilog code that satisfies those same correctness properties specified in Caisson. A strong ability to reason about code properties and to express them in multiple languages is exactly what is called for here, and what diverse LLM test benches help to enable.

In summary, there are good reasons for optimism that automated formal verification will soon be greatly improved. \\

{\bf Acknowledgements:}
The authors wish to thank 
Clark Barrett,
Rustan Leino, 
Daniel Windham,
David Brandfonbrener,
William Byrd,
Josh Engels,
and Anastasiya Kravchuk
for helpful discussions.

\bibliographystyle{unsrtnat}
\bibliography{ref}


\appendix

\section{The Minhash Deduplication Algorithm}
\label{appendixA}

We can think about deduplicating a set of files by finding groups of ``similar''files and then choosing only one file representative from each group to form our final deduplicated set of files. To do this, we can use
 the Jaccard similarity metric to decide whether one document is a duplicate of another. 

The Jaccard similarity metric provides a way to quantify the similarity of two sets. It is defined as \citep{wiki:jaccard_index}:

\[J(A, B) \, = \, \frac{|A \cap B|}{|A \cup B|}\]

In the application to code files, we could consider each file to be a set of $n$-grams, where an $n$-gram is defined as a sequence of $n$ adjacent symbols in a particular order \citep{ngram_wikipedia_2024}, and then apply the Jaccard score as a similarity metric for our files. To directly calculate this Jaccard score, we would need to run string comparison on every $n$-gram, which would have time complexity $O\left(nm^2\right)$ if we have $n$ $n$-grams each with max length $m$ characters. This turns out to be an inefficient method for representing each code file as a set. Instead, the minhash deduplication algorithm approximates the Jaccard similarity between two documents by shingling the documents and comparing the minhash representation of each set of shingles (i.e. we compare fingerprints of documents instead of full documents). The minhash representation of a document is a way to represent a text document as a set of numbers that is faithful to the structure of its content but with a fixed set size that is smaller than the total number of $n$-grams in the document (i.e. the minhash representation of the document is a form of numerical fingerprint of the document). In Figure \ref{fig:minhash} below, we provide the pseudocode for the minhash algorithm used, based entirely on the script in \citep{codededupminhash2024}:

\begin{figure}[htbp]
\begin{lstlisting}
function minhash_deduplication(documents, num_permutations, threshold):
    # Preprocess the documents
    for each document in documents:
        tokenize the document into n-grams (shingles)
        hash each n-gram using a hash function (e.g., xxHash or SHA-1)
        store the hashed n-grams in a set
        
    # Generate permutations
    for i from 1 to num_permutations:
        generate random coefficients a and b
        create a permutation function: (a * x + b) % prime_modulus
        
    # Create minhash signatures
    signatures = []
    for each document in documents:
        signature = []
        for each permutation function:
            min_hash = INFINITY
            for each hashed n-gram in the document:
                permuted_hash = apply permutation function to hashed n-gram
                min_hash = min(min_hash, permuted_hash)
            append min_hash to signature
        append signature to signatures
        
    # Perform Locality-Sensitive Hashing (LSH)
    # We use 250 permutations, so to achieve Jaccard similarity threshold of 0.5
    # We really only need one band (i.e. one hash table)
    num_bands = choose number of bands
    rows_per_band = num_permutations / num_bands
    candidate_pairs = []
    for each band:
        create an empty hash table
        for each document signature:
            band_signature = subset of signature for the current band
            hash_bucket = hash(band_signature)
            add document to the corresponding hash bucket
        for each hash bucket:
            if number of documents in the bucket > 1:
                generate all pairs of documents in the bucket
                add pairs to candidate_pairs
\end{lstlisting}
\caption{Pseudocode for the minhash deduplication algorithm.}
\label{fig:minhash}
\end{figure}

\begin{figure}[htbp]
    \ContinuedFloat
    \begin{lstlisting}
    # Use a union-find datastructure to track groups of duplicates
    duplicates = UnionFind()
    for each band:
        for each row in hashtable:
            for each hash_bucket:
                if size(hash_bucket) <= 1:
                    continue
                else:
                    cluster_id = min(hash_bucket)
                    for x in hash_bucket:
                        duplicates.union(x, cluster_id)
        
    # Perform deduplication
    deduplicated_documents = []
    for each document in documents:
        if duplicates.find_root(document) = document:
            add document to deduplicated_documents
        
    return deduplicated_documents
\end{lstlisting}
\caption{Pseudocode for the minhash deduplication algorithm (continued).}
\end{figure}

Note that the probability two files have the same min hash value under the same hash function is equivalent to their Jaccard similarity. Concretely, for file $A$ and file $B$:

\[\mbox{Pr}\left[\, \min{h_i(A)} = \min{h_i(B)} \,\right] \, = \, J(A, B)\]

where $\min{h_i()}$ denotes taking the minimum hash value under hash function $h_i$. This makes sense because, assuming negligible hash collision, Pr$\, \left[\min{h_i(A)} = \min{h_i(B)}\right]$ is equivalent to the probability that the first $n$-gram hash of $A$ under $h_i$ is equal to the first $n$-gram hash of $B$ under $h_i$. If $h_i$ is a good hash function, then it uniformly distributes the hash values of the original n-gram hashes over the range of $h_i$. Let $c$ denote the number of $n$-grams with equivalent hashes; let $a$ denote the number of $n$-grams from $A$ with smaller hash values than the hash value of corresponding $n$-gram from $B$; let $b$ denote the reverse of the previous category. Then, Pr$\, \left[\min{h_i(A)} = \min{h_i(B)}\right] \, = \, \frac{c}{a + b + c}$, given the uniformity of $h_1$. Note that $\frac{c}{a + b + c} \, = \, \frac{|A \cap B|}{|A \cup B|} \, = \, J(A,B)$.

\section{Prompt Engineering for Hint Reconstruction}
\label{appendixB}

We based our prompts on the prompts used in the \textit{Clover} benchmark \citep{sun2024clover}.

\subsection{GPT Model Famly}

\verb+SYSTEM_PROMPT = "You are an expert in Dafny. You will be given tasks dealing+
\verb+                 with Dafny programs including precise annotations."+

\verb+USER_PROMPT = "Given a Dafny program with function signature, preconditions,+
\verb+               postconditions, and code, but with annotations missing.+
\verb+               Please return a complete Dafny program with the strongest+
\verb+               possible annotations (loop invariants, assert statements,+
\verb+               etc.) filled back in. Do not explain. Please use exactly the+
\verb+               same function signature, preconditions, and postconditions.+
\verb+               Do not ever modify the given lines. Below is the program:"+

\subsection{Claude 3 Opus}
\label{claude_prompts}

\verb+SYSTEM_PROMPT = "You are an expert in Dafny. You will be given tasks dealing+
\verb+                 with Dafny programs including precise annotations. You should+
\verb+                 only return code body in all circumstances. No text is allowed."+

\verb+USER_PROMPT = "Given a Dafny program with function signature, preconditions,+
\verb+               postconditions, and code, but with annotations missing.+
\verb+               Please return a complete Dafny program with the strongest+
\verb+               possible annotation (loop invariants, assert statements,+
\verb+               etc.) filled back in. Do not explain or output any text. If+
\verb+               you have to explain, put all explanations in comments form.+
\verb+               There should only be code body in your output. Please use+
\verb+               exactly the same function signature, preconditions, and+
\verb+               postconditions. Do not ever modify the given lines. Below+
\verb+               is the program:\n```dafny\n"+

\subsection{CodeLlama-7b-Instruct-hf}

The prompts for CodeLlama-7b-Instruct-hf are the same as those in \ref{claude_prompts}.

\section{Proposals for Evaluating Strength of Generated Specifications}
\label{appendixC}

The evaluation of models' capability to generate formal specifications might be enhanced by integrating the process with the creation of positive and negative test cases for each Dafny implementation. This approach proposes a reward system where models are evaluated based on the number of positive test cases their formal specifications support and the number of negative test cases they successfully reject. However, this method introduces a new challenge: ensuring the test cases accurately reflect the comprehensive meaning intended in the natural language descriptions. The consistency and validity of these test cases become critical, raising questions about the methods used to generate and verify them.

\section{Repositories of Scraped Dafny Code}
\label{appendixD}

We provide a full list of all repositories whose data we used in the scraped portion of DafnyBench in Tables \ref{repotable1}, \ref{repotable2}, \ref{repotable3}. When reporting the license information, "Renamed so N/A" implies that the original repository we scraped in December 2023 no longer exists under that name. Otherwise, the repositories have either Microsoft open-source licenses, MIT licenses, GNU General Public License v3.0 licenses, Creative Commons Zero v1.0 Universal, Apache 2.0 licenses, or "Other" (which is secretly an MIT License in a strange format, which has been checked manually). In light of this, we release our derivative DafnyBench repository under an Apache 2.0 license and a GNU General Public License v3.0. We note explicitly here that all files from repositories with the Apache 2.0 license have been modified from their original form.

\begin{table}
    \centering
    \caption{Repositories from which DafnyBench utilizes scraped code (no particular order).\\ \hspace{1cm}}
    \begin{tabular}{ll}
        \toprule
        \textbf{Repository Name} & \textbf{License} \\
        \midrule
        \href{https://github.com/534014913/dafl}{dafl} & No license provided \\
        \href{https://github.com/hath995/Dafny-Grind75}{Dafny-Grind75} & No license provided \\
        \href{https://github.com/vitorhugo13/feup-mfes}{feup-mfes} & MIT License \\
        \href{https://github.com/Eggy115/Dafny}{Dafny} & GNU General Public License v3.0 \\
        \href{https://github.com/wynnliam/nitwit}{nitwit} & MIT License \\
        \href{https://github.com/byd110/Dafny-experiences}{Dafny-experiences} & No license provided \\
        \href{https://github.com/TVSSSRIPAD/Formal_Verification_With_Dafny}{Formal\_Verification\_With\_Dafny} & No license provided \\
        \href{https://github.com/nathand99/SENG2011}{SENG2011} & No license provided \\
        \href{https://github.com/Lystora/M2}{M2} & No license provided \\
        \href{https://github.com/boaz23/assertive-programming-assignment-1}{assertive-programming-assignment-1} & No license provided \\
        \href{https://github.com/notkazz/t1_MF}{t1\_MF} & No license provided \\
        \href{https://github.com/secure-foundations/dafny-exercise}{dafny-exercise} & Other \\
        \href{https://github.com/tareqmahmood/dafny-learn}{dafny-learn} & No license provided \\
        \href{https://github.com/ruipmfs/software-specification-p1}{software-specification-p1} & No license provided \\
        \href{https://github.com/Alex-Amarandei/FMSE-2022-2023}{FMSE-2022-2023} & The Unlicense \\
        \href{https://github.com/Zyfarok/fv2020-tms}{fv2020-tms} & No license provided \\
        \href{https://github.com/SantaC0103/type-definition}{type-definition} & No license provided \\
        \href{https://github.com/hobo0xcc/laboratory}{laboratory} & No license provided \\
        \href{https://github.com/Eggy115/Dafny}{dafny} & GNU General Public License v3.0 \\
        \href{https://github.com/jorge-jbs/TFG}{TFG} & GNU General Public License v3.0 \\
        \href{https://github.com/benreynwar/SiLemma}{SiLemma} & MIT License \\
        \href{https://github.com/Consensys/dafny-training}{dafny-training} & No license provided \\
        \href{https://github.com/LucasGCardoso/FormalMethods}{FormalMethods} & No license provided \\
        \href{https://github.com/longmanxu/dafny_misc}{dafny\_misc} & MIT License \\
        \href{https://github.com/jonhnet/vmware-verification-2023}{vmware-verification-2023} & No license provided \\
        \href{https://github.com/dslogget/CSU55004---Formal-Verification}{CSU55004---Formal-Verification} & No license provided \\
        \href{https://github.com/GambuzX/MIEIC_mfes}{MIEIC\_mfes} & MIT License \\
        \href{https://github.com/vladstejeroiu/Dafny-programs}{Dafny-programs} & No license provided \\
        \href{https://github.com/pemesteves/MFES_2021}{MFES\_2021} & MIT License \\
        \href{https://github.com/Dav1216/DafnyPrograms}{DafnyPrograms} & No license provided \\
        \href{https://github.com/KristienN/cs357}{cs357} & No license provided \\
        \href{https://github.com/IonitaCatalin/formal-methods-in-software-engineering}{formal-methods-in-software-engineering} & No license provided \\
        \href{https://github.com/Sup31/Dafny_ProgrammingLanguages}{Dafny\_ProgrammingLanguages} & No license provided \\
        \href{https://github.com/SteveR-Ncl/CSC8204-Dafny}{CSC8204-Dafny} & No license provided \\
        \href{https://github.com/guimath/BPTree-verif}{BPTree-verif} & No license provided \\
        \href{https://github.com/usrnatc/tangent-finder}{tangent-finder} & No license provided \\
        \href{https://github.com/ArthurSudbrackIbarra/Trab1-Metodos-Formais}{Trab1-Metodos-Formais} & No license provided \\
        \href{https://github.com/chanheec/verified-using-dafny}{verified-using-dafny} & MIT License \\
        \href{https://github.com/ES-PUCRS/Metodos_Formais}{Metodos\_Formais} & No license provided \\
        \href{https://github.com/lemmy/lets-prove-blocking-queue}{lets-prove-blocking-queue} & Creative Commons Zero v1.0 Universal \\
        \href{https://github.com/FanC096/Dafny_Programs}{Dafny\_Programs} & No license provided \\
        \href{https://github.com/Costinteo/dafny-workout}{dafny-workout} & MIT License \\
        \href{https://github.com/AoxueDing/Dafny-Projects}{Dafny-Projects} & No license provided \\
        \href{https://github.com/tannerduve/VerifiedMergeSortDafny}{VerifiedMergeSortDafny} & No license provided \\
        \href{https://github.com/sligocki/dafny_projects}{dafny\_projects} & No license provided \\
        \href{https://github.com/alexiadorneles/pucrs-metodos-formais-t1}{pucrs-metodos-formais-t1} & No license provided \\
        \href{https://github.com/eligoldweber/specTesting}{specTesting} & No license provided \\
        \href{https://github.com/julianafmar/QS_BoilerPlate1}{QS\_BoilerPlate1} & No license provided \\
        \href{https://github.com/namin/dafny-sandbox}{dafny-sandbox} & No license provided \\
        \href{https://github.com/isobelm/formal-verification}{Formal-Verification} & No license provided \\
        \href{https://github.com/Afats/dafny-duck}{dafny-duck} & No license provided \\
        \href{https://github.com/kheirmirza/FlexWeek}{FlexWeek} & No license provided \\
        \href{https://github.com/Aaryan-Patel-2001/703FinalProject}{703FinalProject} & No license provided \\
        \bottomrule
    \end{tabular}
    \label{repotable1}
\end{table}

\begin{table}
    \centering
    \caption{Repositories from which DafnyBench utilizes scraped code (no particular order), continued.\\}
    \begin{tabular}{ll}
        \toprule
        \textbf{Repository Name} & \textbf{License} \\
        \midrule
        \href{https://github.com/pedrovponte/MFS}{MFS} & No license provided \\
        \href{https://github.com/guilhermeolivsilva/dafny-mini-project}{dafny-mini-project} & No license provided \\
        \href{https://github.com/Yrh7383111/Software-Verification}{Software-Verification} & No license provided \\
        \href{https://github.com/soares-eduardo/circular-queue-implemetation}{circular-queue-implemetation} & No license provided \\
        \href{https://github.com/TerrificXu/Final-Project-Dafny}{Final-Project-Dafny} & No license provided \\
        \href{https://github.com/joaopascoalfariafeup/DafnyProjects}{DafnyProjects} & No license provided \\
        \href{https://github.com/minsungc/bbfny}{bbfny} & No license provided \\
        \href{https://github.com/trabajoJorge/Formal-methods-of-software-development}{Formal-methods-of-software-development} & No license provided \\
        \href{https://github.com/BernardoS4/Software-building-and-verification-Projects}{Software-building-and-verification-Projects} & No license provided \\
        \href{https://github.com/Tripparsugo/software_analysis}{software\_analysis} & No license provided \\
        \href{https://github.com/StephRMcIntyre/cs245-verification}{cs245-verification} & No license provided \\
        \href{https://github.com/hath995/dafny-aoc-2019}{dafny-aoc-2019} & No license provided \\
        \href{https://github.com/JoanaSoaresF/ProjectosCVS}{ProjectosCVS} & No license provided \\
        \href{https://github.com/juletx/MFDS}{MFDS} & MIT License \\
        \href{https://github.com/MarkValman/groupTheory}{groupTheory} & No license provided \\
        \href{https://github.com/just-me-/dafny-language-server}{dafny-language-server} & Other \\
        \href{https://github.com/ayush268/Invoker}{Invoker} & Apache License 2.0 \\
        \href{https://github.com/isobelm/formal-verification}{formal-verification} & No license provided \\
        \href{https://github.com/vladstejeroiu/Dafny-programs}{dafny-programs} & No license provided \\
        \href{https://github.com/secure-foundations/ironsync-osdi2023}{ironsync-osdi2023} & Other \\
        \href{https://github.com/dijkstracula/verified-isort}{verified-isort} & No license provided \\
        \href{https://github.com/TonyZhangND/paxos_proof}{paxos\_proof} & No license provided \\
        \href{https://github.com/maddydobbie/se2011}{se2011} & No license provided \\
        \href{https://github.com/jasonthewhale/Dafny_Verify}{Dafny\_Verify} & No license provided \\
        \href{https://github.com/langacristian/Formal-Methods-Project}{Formal-Methods-Project} & No license provided \\
        \href{https://github.com/cassidywaldrip/630-dafny}{630-dafny} & No license provided \\
        \href{https://github.com/monadius/dafny_examples}{dafny\_examples} & MIT License \\
        \href{https://github.com/MicAu/Workshop}{Workshop} & No license provided \\
        \href{https://github.com/cristirusu-99/Dafny-Practice}{Dafny-Practice} & MIT License \\
        \href{https://github.com/DanielCavalheiro/CVS-handout1}{CVS-handout1} & No license provided \\
        \href{https://github.com/tegbesemirone/CS494-final-project}{CS494-final-project} & No license provided \\
        \href{https://github.com/secure-foundations/iron-sync}{iron-sync} & Other \\
        \href{https://github.com/benjaminfjones/stunning-palm-tree}{stunning-palm-tree} & Creative Commons Zero v1.0 Universal \\
        \href{https://github.com/johnterickson/sat_dfy}{sat\_dfy} & No license provided \\
        \href{https://github.com/GLaDOS-Michigan/verification-class}{verification-class} & MIT License \\
        \href{https://github.com/noalero/AssertivePrograming}{AssertivePrograming} & No license provided \\
        \href{https://github.com/dafny-lang/Dafny-VMC}{Dafny-VMC} & MIT License \\
        \href{https://github.com/dafny-lang/libraries}{libraries} & Other \\
        \href{https://github.com/lamula21/cmsc433}{cmsc433} & No license provided \\
        \href{https://github.com/FaizAther/Correctness}{Correctness} & No license provided \\
        \href{https://github.com/VicentF/CVS-Projto1}{CVS-Projto1} & No license provided \\
        \href{https://github.com/Nangos/dafleet}{dafleet} & MIT License \\
        \href{https://github.com/SwampertX/dafny-rope}{dafny-rope} & MIT License \\
        \href{https://github.com/tchajed/protocol-verification-fa2023}{protocol-verification-fa2023} & No license provided \\
        \href{https://github.com/olrodr03/vfag}{vfag} & No license provided \\
        \href{https://github.com/PaddyZz/Dafny_Learning_Experience}{Dafny\_Learning\_Experience} & Apache License 2.0 \\
        \href{https://github.com/wenhuizhang/summer-school-2020}{summer-school-2020} & No license provided \\
        \href{ }{BinarySearchTree} & Renamed so N/A  \\
        \href{ https://github.com/namin/llm-verified-with-monte-carlo-tree-search}{llm-verified-eval} &  MIT License \\
        \href{ }{Programmverifikation-und-synthese} & Renamed so N/A  \\
        \href{ }{Prog-Fun-Solutions} & Renamed so N/A  \\
        \href{ }{CO3408-Advanced-Software-Modelling-Assignment...} & Renamed so N/A \\
        \bottomrule
    \end{tabular}
    \label{repotable2}
\end{table}

\begin{table}[ht!]
    \centering
    \caption{Repositories from which DafnyBench utilizes scraped code (no particular order), continued.\\}
    \begin{tabular}{ll}
        \toprule
        \textbf{Repository Name} & \textbf{License} \\
        \midrule
        \href{https://github.com/Caitlin-Goodger/DafnyExercises}{DafnyExercises} & No license provided \\
        \href{https://github.com/byu-dafny/test-generation-examples}{test-generation-examples} & No license provided \\
        \href{https://github.com/dakotawong/HATRA-2022-Paper}{HATRA-2022-Paper} & No license provided \\
        \href{https://github.com/volodeyka/veri-sparse}{veri-sparse} & No license provided \\
        \href{https://github.com/Rayyan-Mehmood/Formal-Verification-Project}{Formal-Verification-Project} & No license provided \\
        \href{https://github.com/j-kanbour/formal_verication_dafny}{formal\_verication\_dafny} & No license provided \\
        \href{https://github.com/priyadudhe/Simulink-To_dafny}{Simulink-To\_dafny} & No license provided \\
        \href{https://github.com/benreynwar/dafny_experiments}{dafny\_experiments} & No license provided \\
        \href{https://github.com/dwebb9/cs686}{cs686} & No license provided \\
        \href{https://github.com/kyrolloszakaria/Program-Verification-Dataset}{Program-Verification-Dataset} & MIT License \\
        \href{https://github.com/Flavius88/Dafny-demo}{Dafny-demo} & No license provided \\
        \href{https://github.com/zhuzilin/dafny-exercises}{dafny-exercises} & No license provided \\
        \href{https://github.com/tecnicasilegais/MetodosFormais}{metodosFormais} & No license provided \\
        \href{https://github.com/tatayu/CS5232_Project}{CS5232\_Project} & No license provided \\
        \href{https://github.com/zhuzilin/dafny-exercises}{Dafny-Exercises} & No license provided \\
        \bottomrule \\
    \end{tabular}
    \label{repotable3}
\end{table}


\section{Dafny Verification Examples}
\label{verification-examples}

We take one example test program from DafnyBench, and consider four possible results for the corresponding LLM-reconstructed program: successfully verifies, fails to verify, cheats by including \texttt{assume false}, and cheats by including \texttt{\{:verify false\}}. The last three cases are all considered a fail by the DafnyBench evaluation metric.

\subsection{Successful Example}

Figure \ref{fig:successful_example} shows a Dafny program that is considered to have successfully verified without cheating.

\begin{figure}[ht!]
\begin{lstlisting}
function sorted(a: array<int>) : bool
    reads a
{
   forall i,j : int :: 0 <= i < j < a.Length ==> a[i] <= a[j]
}

method BinarySearch(a: array<int>, x: int) returns (index: int)
    requires sorted(a)
    ensures 0 <= index < a.Length ==> a[index] == x
    ensures index == -1 ==> forall i : int :: 0 <= i < a.Length ==> a[i] != x
{
    var low := 0;
    var high := a.Length - 1;
    var mid := 0;
    
    while (low <= high) 
        invariant 0 <= low <= high + 1 <= a.Length
        invariant x !in a[..low] && x !in a[high + 1..]
    {
        mid := (high + low) / 2;
        if a[mid] < x {
            low := mid + 1;
        }
        else if a[mid] > x {
            high := mid - 1;
        }
        else {
            return mid;
        }
    }
    return -1;
}
\end{lstlisting}
\caption{An example response that successfully fills hints back in and verifies without cheating.}
\label{fig:successful_example}
\end{figure}

\textbf{Dafny verifier message}: Dafny program verifier finished with 3 verified, 0 errors.

\subsection{Failed Example}

Figure \ref{fig:failed_example} shows a Dafny program that fails to be verified.

\begin{figure}[ht!]
\begin{lstlisting}
function sorted(a: array<int>) : bool
    reads a
{
   forall i,j : int :: 0 <= i < j < a.Length ==> a[i] <= a[j]
}

method BinarySearch(a: array<int>, x: int) returns (index: int)
    requires sorted(a)
    ensures 0 <= index < a.Length ==> a[index] == x
    ensures index == -1 ==> forall i : int :: 0 <= i < a.Length ==> a[i] != x
{
    var low := 0;
    var high := a.Length - 1;
    var mid := 0;
    
    while (low <= high) 
    {
        mid := (high + low) / 2;
        if a[mid] < x {
            low := mid + 1;
        }
        else if a[mid] > x {
            high := mid - 1;
        }
        else {
            return mid;
        }
    }
    return -1;
}
\end{lstlisting}
\caption{An example response that fails to be verified, as it lacks necessary \texttt{invariant} statements.}
\label{fig:failed_example}
\end{figure}

\textbf{Dafny verifier message}: (20,11): Error: index out of range. (30,4): Error: a postcondition could not be proved on this return path. (11,28): Related location: this is the postcondition that could not be proved. Dafny program verifier finished with 2 verified, 2 errors.

\subsection{Cheat Example}

Figure \ref{fig:cheat_example_1} shows that a Dafny program cheats by including \verb+assume false+, which DafnyBench evaluation would count as a fail.

\begin{figure}[ht!]
\begin{lstlisting}
function sorted(a: array<int>) : bool
    reads a
{
   forall i,j : int :: 0 <= i < j < a.Length ==> a[i] <= a[j]
}

method BinarySearch(a: array<int>, x: int) returns (index: int)
    requires sorted(a)
    ensures 0 <= index < a.Length ==> a[index] == x
    ensures index == -1 ==> forall i : int :: 0 <= i < a.Length ==> a[i] != x
{
    assume false;
    var low := 0;
    var high := a.Length - 1;
    var mid := 0;
    
    while (low <= high) 
    {
        mid := (high + low) / 2;
        if a[mid] < x {
            low := mid + 1;
        }
        else if a[mid] > x {
            high := mid - 1;
        }
        else {
            return mid;
        }
    }
    return -1;
}
\end{lstlisting}
\caption{An example response that cheats by including \texttt{assume false}.}
\label{fig:cheat_example_1}
\end{figure}

\textbf{Dafny verifier message}: Dafny program verifier finished with 3 verified, 0 errors.

\subsection{Another Cheat Example}

Figure \ref{fig:cheat_example_2} shows that another Dafny program cheats by including \texttt{\{:verify false\}}, which DafnyBench evaluation would count as a fail.

\begin{figure}[ht!]
\begin{lstlisting}
function sorted(a: array<int>) : bool
    reads a
{
   forall i,j : int :: 0 <= i < j < a.Length ==> a[i] <= a[j]
}

method {:verify false} BinarySearch(a: array<int>, x: int) returns (index: int)
    requires sorted(a)
    ensures 0 <= index < a.Length ==> a[index] == x
    ensures index == -1 ==> forall i : int :: 0 <= i < a.Length ==> a[i] != x
{
    var low := 0;
    var high := a.Length - 1;
    var mid := 0;
    
    while (low <= high) 
    {
        mid := (high + low) / 2;
        if a[mid] < x {
            low := mid + 1;
        }
        else if a[mid] > x {
            high := mid - 1;
        }
        else {
            return mid;
        }
    }
    return -1;
}
\end{lstlisting}
\caption{An example response that cheats by including \texttt{\{:verify false\}}.}
\label{fig:cheat_example_2}
\end{figure}

\textbf{Dafny verifier message}: Dafny program verifier finished with 3 verified, 0 errors.

\section{Overdetailed Specification}
\label{over_detailed_spec}

Figures \ref{fig:clover_example_1} and \ref{fig:clover_example_2} show two example programs \texttt{update\_array\_strong.dfy} and \texttt{triple\_strong.dfy} from the \textit{Clover} benchmark \citep{sun2024clover}, in which the formal specification closely echoes the program implementation.

\begin{figure}[ht!]
\begin{lstlisting}
method UpdateElements(a: array<int>)
  requires a.Length >= 8
  modifies a
  ensures old(a[4]) +3 == a[4]
  ensures a[7]==516
  ensures forall i::0 <= i<a.Length ==> i != 7 && i != 4 ==> a[i] == old(a[i])
{
  a[4] := a[4] + 3;
  a[7] := 516;
}
\end{lstlisting}
\caption{An example program \texttt{update\_array\_strong.dfy} from the \textit{Clover} benchmark \citep{sun2024clover}, in which the formal specification closely echoes the program implementation.}
\label{fig:clover_example_1}
\end{figure}

\begin{figure}[ht!]
\begin{lstlisting}
method Triple (x:int) returns (r:int)
  ensures r==3*x
{
  r:= x*3;
}
\end{lstlisting}
\caption{Another example program \texttt{triple\_strong.dfy} from the \textit{Clover} benchmark \citep{sun2024clover}, in which the formal specification closely echoes the program implementation.}
\label{fig:clover_example_2}
\end{figure}

\section{Ethics Statement}
In creating DafnyBench, we took care to use only data that was publicly available on GitHub, and we reference every repository from which we acquired this data, along with their licenses, in Appendix \ref{appendixD}. Furthermore, we cite the existing verifiable programming benchmarks that we subsume in DafnyBench (i.e. {\it  Clover} \cite{sun2024clover} and \textit{dafny-synthesis} \cite{MRHMisuDafnyFSE24}), and we asked explicit permission from their authors in order to do so. Finally, we cite all models that were used for evaluations on this benchmark \cite{openai_2023_gpt4, brown2020gpt3, claude_3_tech_report, huggingface2022llama}. We used these models in accordance with the policies set forth in their API and model card documentation.

\section{Reproducibility Statement}
\label{reproducability}
Our benchmark contains the 782 \texttt{ground\_truth} programs and the corresponding \texttt{hints\_removed} programs. Additionally, we include full metadata on all of these files and the evaluation scripts necessary for running the listed models on them. By using the OpenAI and Anthropic APIs, others looking to reproduce this work should not expect to spend more than $\$300$ for a full run of GPT4-o on DafnyBench, $\$300$ for a full run of Claude3 on DafnyBench, $\$500$ for a full run of GPT4-turbo on DafnyBench, and $\$400$ for a full run of GPT-3.5 on DafnyBench. We used the \texttt{sglang} package \cite{zheng2023efficiently} to efficiently query the models. All evaluations were completed on a Linux cluster with an A100 Nvidia GPU.

\end{document}